\documentstyle[floats,multicol,aps,epsfig,prl,subfigure]{revtex}

\newlength{\bxwidth}\bxwidth=3.2 truein

\newlength{\jight}
\newlength{\fight}
\newcommand{\fg}[4]
{\begin{figure}[h]\epsfxsize=#1
\centerline{\epsfbox{#2}}\vskip 0.1truein
\caption{{#3}}\label{#4}\end{figure}}
\newcommand\ltdash{\raise-1.8pt\hbox{$\scriptscriptstyle |$}}
\newcommand \beq  {\begin{equation}}
\newcommand \eeq  {\end{equation}}
\newcommand \bea {\begin{eqnarray} }
\newcommand \eea {\end{eqnarray}}

\newcommand \ra { \rangle}
\newcommand\la{\langle}

\newcommand\etal{{\it et al.}}

\begin{document}
\draft
\twocolumn[\hsize\textwidth\columnwidth\hsize\csname @twocolumnfalse\endcsname
\title{
The fate of Kondo resonances in certain Kondo lattices: a ``poor
woman's'' scaling analysis}
\author{ Catherine P\'epin }
\address{ SPhT, L'Orme des Merisiers, CEA-Saclay, 91191 Gif-sur-Yvette, France
}

\maketitle
\date{\today}
\maketitle
\begin{abstract}

We present an effective field theory for the Kondo lattice, which
can exhibit, in a certain range of parameters, a non Fermi liquid
paramagnetic phase at the brink of a zero temperature Anti
Ferromagnetic (AF) transition. The model is derived in a natural
way from the bosonic Kondo-Heisenberg model, in which the Kondo
resonances are seen as true (but damped) Grassmann fields in the
field theory sense. One loop Renormalization Group (RG) treatment
of this model gives a phase diagram for the Kondo lattice as a
function of $ J_K  $ where, for $ J_K< J_c$ the system shows AF
order, for $J_K> J_1$ one has the heavy electron phase and for $
J_c < J_K < J_1$ the formation of the Kondo singlets is
incomplete, leading to the breakdown of the Landau Fermi liquid
theory.

\end{abstract}
\vskip 0.2 truein 
\newpage
\vskip2pc]

In the last ten years, there has been an increasing body of
experimental evidence showing striking deviations from
conventional Landau Fermi Liquid theory (LFL) at some heavy
fermion quantum critical points (QCP)
\cite{review,lohneysen,sido,mathur,grosche,julian,aoki,aeppli,steglich,gegenwart}.
The specific heat coefficient is seen to diverge at the QCP,
showing a logarithmic increase
\cite{lohneysen,aoki,steglich,aeppli} as the temperature is
decreased, followed for two compounds (YbRh$_2$Si$_2$ and
CeCoIn$_5$ tuned to criticality under magnetic field), by an
upturn at lower temperature \cite{steglich,gegenwart,movshovich}.
The resistivity is quasi-linear in temperature for most compounds
\cite{grosche,gegenwart,julian}, with perfectly linear dependence
for YbRh$_2$(Si$_{0.95}$Ge$_{0.05}$)$_2$ and
CeCu$_{5.9}$Au$_{0.1}$ \cite{review,lohneysen,steglich}. NMR and
$\mu$-SR studies show that, generically, the susceptibility
acquires anomalous exponents \cite{grosche,gegenwart,aeppli}. For
CeCu$_{6-x}$Au$_x$ \cite{aeppli} neutron scattering measurements
reveal the presence of $B / T$ and $\omega / T$ scaling in the
dynamic spin susceptibility at the QCP. Recent e-SR~\cite{esr},
thermal expansion~\cite{kuchler} as well as Hall
effect~\cite{silke,matsuda}, heat transport~\cite{taillefer} and
Nernst effect~\cite{kamran} studies show further deviation from
the LFL predictions. Last, for both the compounds
CeCoIn$_5$~\cite{paglione} and YbRh$_2$(Si$_{0.95}$
Ge$_{0.05}$)$_2$~\cite{jeroenphd} driven to criticality by
applying a magnetic field, there are indications that deviations
from the LFL theory of metals might appear over a whole region
--instead of a point-- of the phase diagram, at the vicinity of
the QCP.

 These
very striking results have inspired several theoretical
descriptions. Some require fully anisotropic (2D) spin
fluctuations \cite{2dsdw,si}. Recently the idea of a local mode at
criticality\cite{review,aeppli,si} has emerged in relation to the
compounds CeCu$_{6-x}$Au$_x$ \cite{aeppli} and YbRh$_2$Si$_2$
\cite{steglich,gegenwart}. There are also some approaches invoking
deconfinement and ``fractionalization'' in gauge theories
\cite{senthil}, but none of the theoretical approaches so far, can
account for a non Fermi liquid paramagnetic phase, as well as can
fit more than one or two experimental observations.


In this paper we present an effective model for the Kondo lattice,
which exhibits a non Fermi liquid paramagnetic phase at the
vicinity to a zero temperature AF transition or QCP. The starting
point is the Kondo-Heisenberg lattice model, where we use a
Schwinger boson representation for the spins of the impurities. We
call this model Bosonic Kondo-Heisenberg model (BKH). In this
framework, the Kondo bound states are represented by spinless
Grassmann fields ($ \chi^\dagger $, $\chi$), which have no
expectation value at the mean-field level. On the other hand, the
bosonic representation of the impurity spins enables us to get a
decent treatment of antiferromagnetism at the mean-field level. We
observe that, starting from high energy and going to the infra-red
sector, the Kondo bound states ($\chi^\dagger$, $\chi$) acquire
dynamics, damping and dispersion, since they are coupled to
"itinerant" excitations. The key ingredient of our effective
theory is thus to treat the Grassmann Kondo resonances as {\it
true fields} in the field theory sense. We attribute to them some
dynamics as well as some damping from the start, and couple them
in a physical way to fields of the BKH model. A Renormalization
Group (RG) treatment applied to this model has the remarkable
property that, both the formation of Kondo singlets and the AF
fluctuations -- which prevent Kondo singlets to be formed-- appear
at the one loop level in the form of a logarithmic singularity.
That the formation of Kondo singlets has a logarithmic signature
at the one loop level is a well known fact, in the context of the
Kondo impurity, where it has been observed a long time
ago~\cite{anderson,nozieres}. The new feature, here, is the
presence of a logarithm at the one loop, characteristic of the AF
fluctuations. This is the direct consequence of treating the
resonances as true fields, with some intrinsic dynamics as well as
damping. We are then in a position to obtain a phase diagram for
the Kondo lattice, which can exhibits non Fermi liquid properties
over a finite region of the diagram.


The BKH lattice Hamiltonian \bea \label{BKH}
H  =   H_c  + & H_K & + H_H \  ,  \nonumber \\
\mbox{where} \ H_c & = & \sum_{k \sigma} \varepsilon_k
f^\dagger_{k \sigma} f_{k
\sigma} \ ,  \nonumber \\
H_K & =&  J_K \sum_{i \sigma \sigma^\prime} b^\dagger_{i \sigma}
b_{i \sigma^\prime} f^\dagger_{i \sigma^\prime} f_{ i \sigma} \ , \nonumber  \\
H_ H & = &  J_H \sum_{(i,j) \sigma \sigma^\prime} b^\dagger_{i
\sigma} b_{i \sigma^\prime} b^\dagger_{j \sigma^\prime} b_{j
\sigma} \eea describes the conduction band ($H_c$), the Kondo
coupling between local moments and the conduction electrons at
site $i$ ($H_K$), and the super-exchange between neighboring spins
($H_H$).

The physics of this model depends on the ratio $x =J_K/ J_H$. In
the Doniach scenario~\cite{doniach} for the Kondo lattice, when $x
\ll 1$ the spins anti-ferromagnetically order, and as $x$ is
increased the AF state undergoes a transition towards a heavy
electron paramagnet. It is the goal of this paper to take a closer
look at the nature of this transition.

When we formulate the BKH model as a functional integral, we can
decouple the fields as follows \bea H_K & \rightarrow & H^\prime_K
= \sum_{i \sigma} \left [ b^\dagger_{i \sigma} \chi_i^\dagger f_{i
\sigma} + h.c. \right ] -
\frac{\chi^\dagger_i \chi_i}{J_K} \\
H_H & \rightarrow & H^\prime_H =\sum_{(i,j) \sigma} \left [
|\Delta_{ij}| e^{i \frac{\pi}{a} {\bf r_i- r_j}} b^\dagger_{i
\sigma} b_{j -\sigma}^\dagger + h.c. \right] - \frac{|
\Delta_{ij}|^2}{J_H} \ , \nonumber \eea where the bond variable in
the first term is a Grassmann field which doesn't carry a spin and
the bond variable in the second term has been chosen following an
SP(2N) decomposition of the interaction~\cite{sachdev-spn}. The
mean-field theory of this model requires an additional constraint
term on the Schwinger boson~\cite{oldpiers,olivier} representation
of the impurity spin
\[ H = H_c + H_K + H_H + \sum_i \lambda \left ( b^\dagger_{i
\sigma} b_{i \sigma} - 2 S \right ) \ , \] where $S$ is the spin
of the impurity, taken to be $ 1/2$ in the case of our concern.

At the mean-field level, the $\chi$-fermions don't acquire any
expectation value, while the bosonic bound state $\Delta$ can
condense, leading to an AF phase. After diagonalizing the bosonic
part of $H$ in a mean-field approximation, one gets the following
propagator for the spinons \bea \label{mfieldAF} {\hat G( \omega,
k)} & = & \left [
\begin{array}{cc} \la T b_{k \sigma} b^\dagger_{-k -\sigma} \ra &
\la T b_{k \sigma} b_{-k -
\sigma} \ra \\
\la T b^\dagger_{ -k -\sigma} b_{k \sigma}^\dagger \ra & \la T
b^\dagger_{-k -\sigma} b_{-k -\sigma} \ra \end{array} \right]
\nonumber
\\
& = & \frac{1}{ (i \omega)^2 - \omega_k^2} \left [
\begin{array}{cc}
i \omega - \lambda & \Delta_k \\
\Delta_k & - i \omega - \lambda \end{array} \right ] \ , \eea
where $ \omega$ is a bosonic Matsubara frequency, $ \Delta_k =
\Delta \left ( \sin k_x + \sin k_y + \sin k_z \right )$ is an odd
function of $k$ and $ \omega_k = \sqrt{ \lambda^2 - \Delta_k^2}$.
If $ |\lambda| > \Delta $ the system is in the paramagnetic phase
and spinons have a mass $ \omega_0 = \sqrt{ \lambda^2 -
\Delta^2}$. When $ |\lambda| = \Delta $ the system is at a second
order phase transition towards an AF ground state and spinons
become massless ($ \omega_0 = 0$). In what follows we consider
that the Kondo lattice undergoes an AF transition, such that part
of the phase diagram is paramagnetic ( $ \omega_0 \neq 0 $) and
part is antiferromagnetic ($ \omega_0 =0$). $\omega_0$ is the mass
of the spinons and plays the role of a parameter in the effective
theory.


Through coupling to the spinons ( $b_{k \sigma}$) and the
itinerant electrons ($f_{k \sigma}$) the $\chi$-fermion Kondo
resonances acquire some dynamics, damping and dispersion. It is
natural to think that at some energy scale the dynamics will
become linear in $\omega$ --actually a propagator linear in
$\omega$ is needed to regulate the high energy dependence. We
propose the following effective action, comprising three fields
\bea \label{model} S &= &S_f + S_b + S_\chi + S_{int} \ , \\
S_f & = & \int d \omega d^dk \sum_\sigma f^\dagger_{k \sigma}
\left (
i \omega - \varepsilon_k \right ) f_{k \sigma} \ , \nonumber \\
S_b & = &\int d \nu d^d q \sum_\sigma V_b^\dagger {\hat G}^{-1}
(\nu, q) V_b \ , \nonumber \\
S_\chi & = & \int d \omega d^d p \chi^\dagger_p \left ( i \omega +
i
\Gamma (\omega ) sgn( \omega) + g^2_0/ J_K \right ) \chi_p  \ ,\nonumber \\
S_{int} & = & g \int (d \omega)^2  d^d k d^dq d^d p \delta_{\bf p
+ q -k} \sum_\sigma \left ( b^\dagger_{q \sigma} \chi^\dagger_p
f_{k \sigma} + h. c. \right ) \nonumber , \eea where $ V_b = \left
( \begin{array}{l} b_{q \sigma} \\ b^\dagger_{-q -\sigma}
\end{array} \right ) $ is the spinor basis for the mean-field
description (\ref{mfieldAF}) of the AF fluctuations, $ \Gamma (
\omega )$ is the damping associated with the $\chi$-fermions,
which we take constant in order to simplify computations, but
which in all generality is taken to be relevant in the infrared
sector ( for example $ \Gamma ( \omega ) = |\omega|^\alpha $ with
$ \alpha < 1 $), and $g$ is the interaction between the three
fields. We have re-scaled the $\chi$ fields such that their
inverse propagator has the dimension of energy.  Now the
$\chi$-fermions are considered as true fields with a finite
lifetime  and a dispersion $\varepsilon_\chi (p) = - g^2_0 / J_K
$. The dispersion becomes a tuning parameter of our theory.
Generically, three cases are possible:
    (1) The $\chi$-``band'' is full, or $J_K >0$. This case
    corresponds to AF Kondo coupling and is considered in this
    paper. The dispersion of the band is neglected as long as the
    band is full.
    (2) The $\chi$-``band'' is empty, or $J_K <0$. This is the
    ferromagnetic Kondo coupling, where the interaction $g$ is
    marginally irrelevant. This case is of no great interest for
    HF compounds, but can be of some relevance for the manganites.
    (3) The $\chi$-``band'' is
    so much renormalized through its interaction with the spinons
    and the conduction electron, that it acquires a ``Fermi''
    surface--or a surface in $k$-space where the dispersion
    changes
    sign. This is a very intriguing possibility, allowing the
    presence of {\it massless} $\chi$-fermions. This case
    will be addressed in a future work.


We now turn to the study of the stability of the model with
respect to the coupling $g$ (diagram c in Fig \ref{vertices}). One
peculiarity of our model is that at one loop, there is no
renormalization of the vertex coming from irreducible diagrams.
The logarithmic contribution comes from the renormalization of the
spinon and $\chi$-fermion propagators (diagrams a and b in figure
\ref{vertices}).
 \fg{\bxwidth}{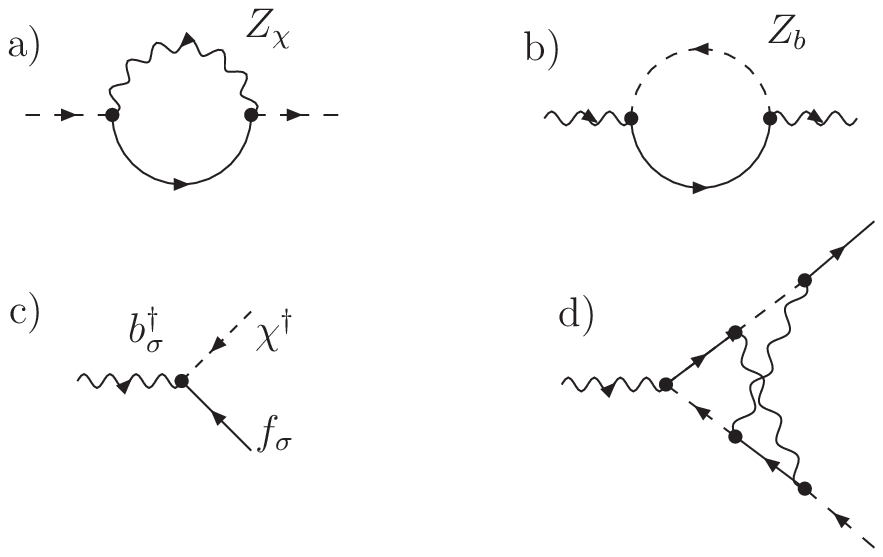}{Diagrams associated
with a) the renormalization of the $\chi$-fermion field, b) the
spinon field c) the vertex $g$ and d) two loops renormalization of
the vertex.
  Solid, dashed and wavy lines are respectively $f$-fermion, $\chi$-fermion  and boson
propagators.}{vertices} We use a RG scheme, where the parameter
$g_0^2/ J_K$ as well as $ \lambda$ are fixed, while the
counter-terms are absorbed into the renormalization of the fields
$ \chi = Z_{\chi}^{1/2}  \chi^R$ and $b = Z_b^{1/2}  b^R$, where
$\chi^R$ and $b^R$ are the renormalized fields. The invariance of
the propagators is ensured by introducing  two new parameters,
keeping track of the field's lifetime $ G^{-1}_\chi = Z_{\chi
\omega}  i \omega   + g^2_0 / J_K $ and $ G_b^{-1} = Z_{b \omega}
i \omega - \lambda $. The $\beta$-function for the local
interaction $g$ takes the form
 \beq \label{oneloop}
 \beta(g) = \frac{d g}{d \log D} = -  \pi \rho g^3 \left ( J_K / g_0^2 -
 1/ \lambda \right ) \ , \eeq where $D$
 is the bandwidth of the conduction electrons, $\rho$ is the
 density of states and $\lambda$ is generically
 positive, such that for $J_H =0$ in (\ref{BKH}) spinons have
 poles at positive energies.
 In the flow (\ref{oneloop}) the first term comes from the
 renormalization of the Kondo resonances (diagram a in Fig.
 \ref{vertices}) and is responsible for the usual Kondo behavior
 of the model, leading $g$ to strong coupling.
 The new feature is the presence of a logarithm at
 one loop, coming from the renormalization of the spinon
 propagator (diagram b in Fig. \ref{vertices}) and opposing the
 Kondo effect in the RG flow. Note that this term wouldn't be present if
 we didn't give {\it dynamics} to the $\chi$-fields in (\ref{model}).
\begin{figure}
\epsfig{figure=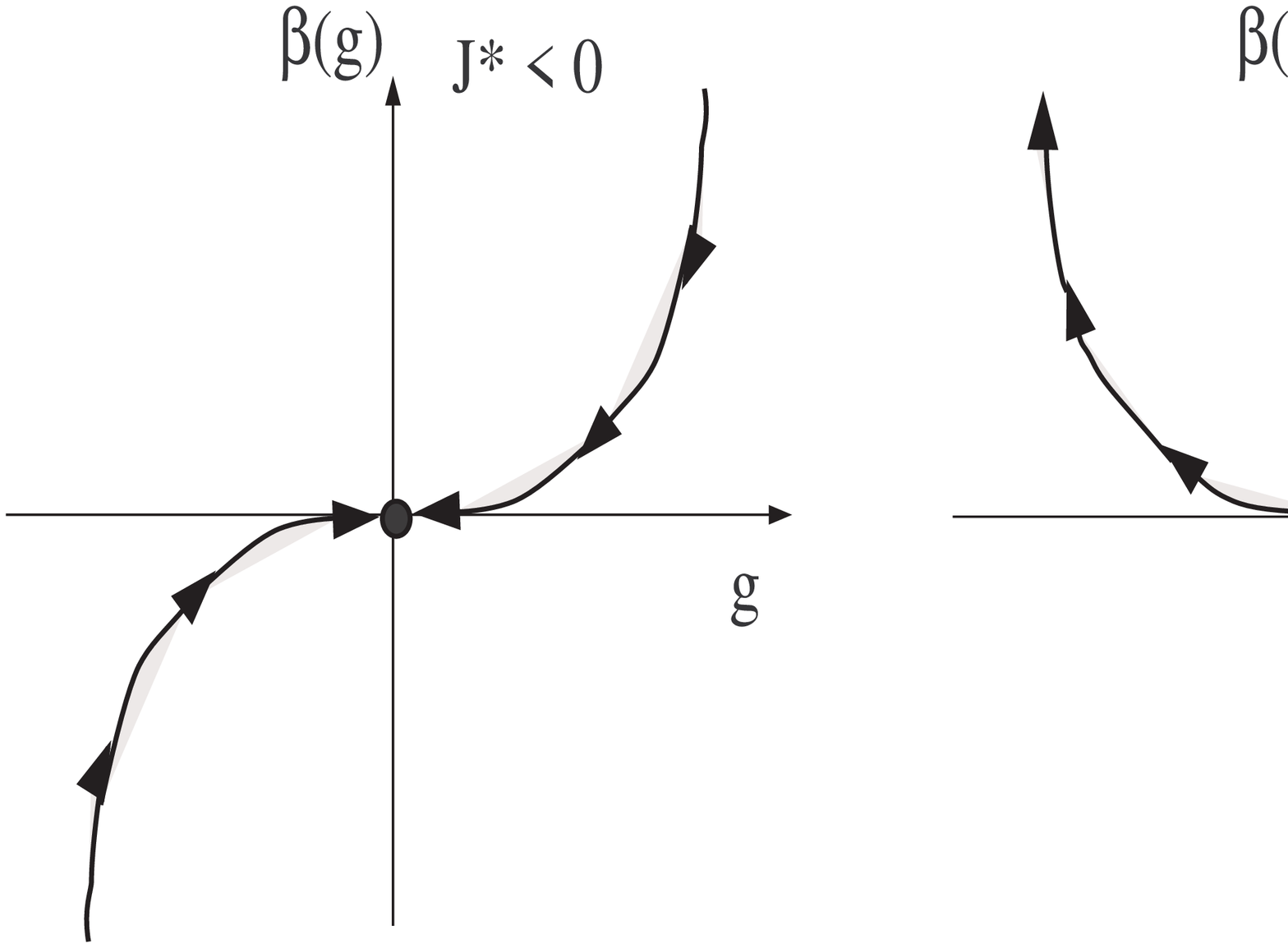,width=6 cm} \caption{ RG flows for the
coupling constant $g$.
 }  \label{flows}
\end{figure}
Two different possible flows are shown in Fig.\ref{flows}. For
 $J_K - g_0^2 / \lambda > 0$, the flow goes
 to strong coupling. We identify the strong coupling fixed
 point with the heavy Fermi liquid phase, where Kondo singlets form.
 On the other hand, for $J_K  - g_0^2 / \lambda < 0$,
 the flow goes to weak coupling and the model is stable. Note that
 the stability criterion doesn't depend on the sign of $g$.

Integrating (\ref{oneloop}) enables to get the energy scales:
\[
\begin{array}{cc}\displaystyle{
g^2 = \frac{g_0^2}{ 1 + 2 \pi \rho  J^* \log(T/D)} \ ;}&
\displaystyle{ T^* = D \exp \left [-1/ (2 \pi \rho J^*) \right ] \
,}
\end{array}
 \] where $J^* =J_K - g_0^2 / \lambda $,  and $T^*$
 is the temperature at which the Kondo singularity appears, or in
 other words the ``lattice'' Kondo temperature.
 One notices, that compared to the single impurity $T_K = D \exp
 \left [ -1 / ( 2 \pi \rho J_K ) \right]$, the lattice Kondo
 temperature $T^*$ is reduced by the presence of AF
 fluctuations.

 Examination of the field's lifetimes leads to
 \beq \begin{array}{cc}
 {\displaystyle \frac{d Z_{\chi \omega}}{ d \log D} = - 2 \pi \rho J_K g^2 / g_0^2 } \
 ; & {\displaystyle \frac{d Z_{b \omega}}{ d \log D } = 2 \pi \rho g^2 /
 \lambda }
 \end{array}\eeq The lifetime of the $\chi$-fermion becomes
 infinitely small in the IR sector, which signals a break down of the Landau Fermi
 liquid theory, while at one loop the spinon lifetime is stable.

 Three phase diagrams are possible, depending on the strength of
 the AF fluctuations.  At large $J_K$ we are, at zero temperature, in a
 heavy Fermi liquid phase, where our model flows to strong
 coupling. The first possibility is that $J^*$ stays {\it
 positive} in the whole paramagnetic phase, reaching the AF QCP
 while the heavy quasi-particles are still well formed. In this
 scenario, conduction electrons are destabilized by a Spin Density
 Wave (SDW)~\cite{moriya} connecting portions of the
 Fermi surface. The best candidate for this scenario is surely
 CeNi$_2$Ge$_2$, for which a study of the Gr\"uneisen parameter has shown
 reasonable agreement with 3D Moriya theory~\cite{kuchler}.

 The second possibility is that $J^*$ stays positive in the whole
 paramagnetic phase, but vanishes at the AF QCP, where the
 fluctuations are the strongest. This happens for $ J_K = g_0^2 /
 \lambda_{min} =g_0^2 / \Delta = g_0^2/ (S J_H)$, in a large $S$ expansion of the
 Heisenberg model.

\begin{figure}
\epsfig{figure=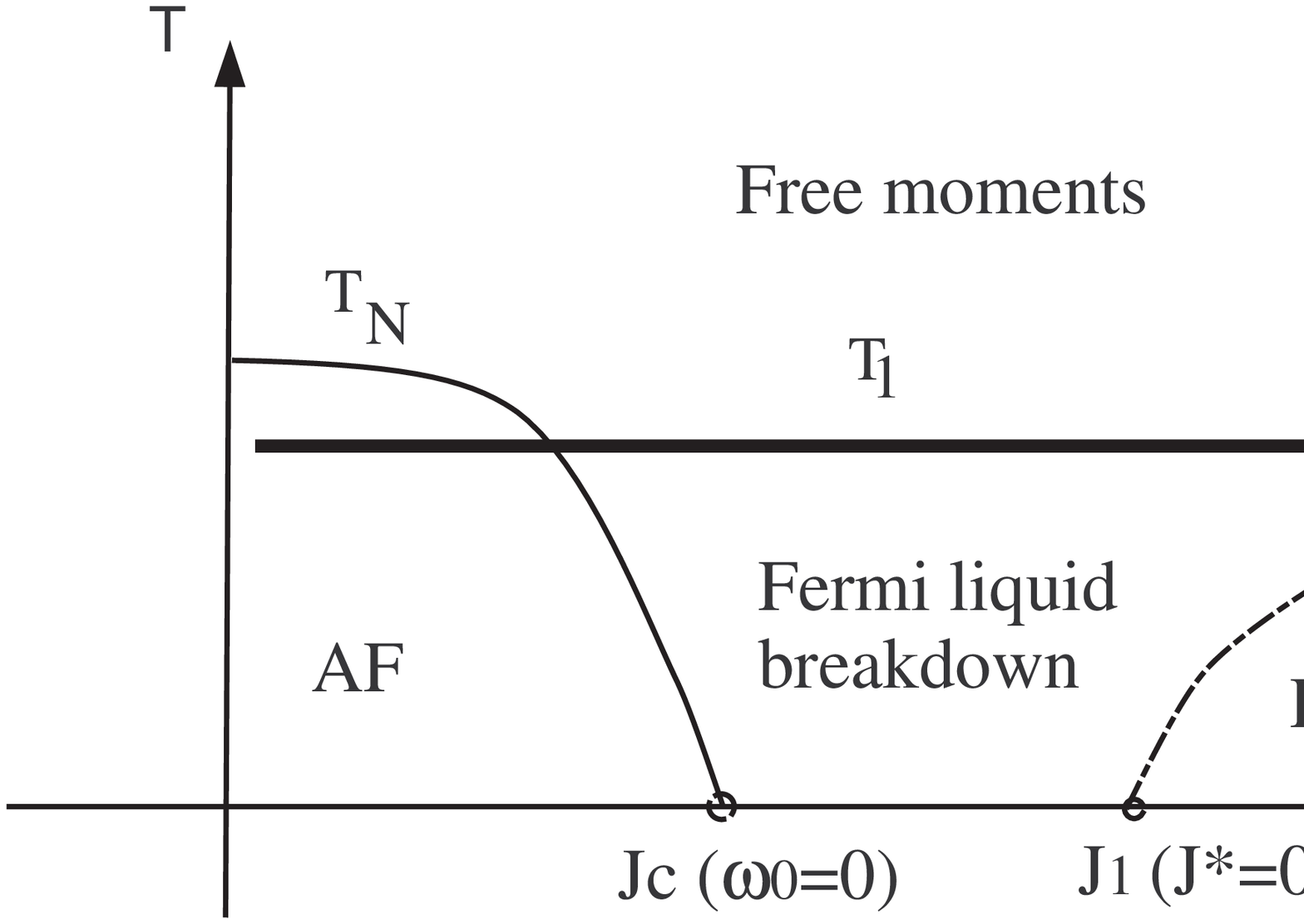,width=6 cm} \caption{ Phase diagram for
$J_K < g_0^2 /\Delta$} \label{phase_diagram}
\end{figure}
 Last, it might happen that the AF fluctuations close to the QCP
 are so strong that the Kondo effect is broken already in the
 paramagnetic phase, when $J^*$ changes sign {\it before} reaching
 the AF transition
 (see Fig.\ref{phase_diagram}). This happens for $J_K < g_0^2 /\Delta$.
  In this case, the Landau Fermi liquid
  theory is broken in a whole {\it region} of the phase
  diagram for which our model (\ref{model}) is stable.
  Recently, measurements on CeCoIn$_5$~\cite{kamran,paglione} and
 YbRh$_2$(Si$_{9.95}$ Ge$_{0.05}$)$_2$~\cite{jeroenphd} indicate the presence of such a
 phase.

In order, to further justify our identification of the strong
 coupling with the heavy Fermi liquid phase, we derived the RG
 flow at two loops. The diagram d in figure \ref{vertices} is the
 sole contributor, and we find
 \beq
 \beta (g) = - 2 \pi \rho g^3 \left ( J^* +  2 \pi \rho g^2 A
 \log(g^2/J_K) \right ) \ , \eeq where
 $A= 8 \pi \lambda \left ( \omega_m - \omega_0) / v_s \right )^2$
 in $D=3$,
 with $\omega_m$ and $\omega_0$ the upper and lower energies of the spinons and
 $v_s$ the spinon dispersion $\omega_q = v_s q$ taken linear over the whole spectrum.
 Since at high energy the starting
 parameters of our model are $g_0^2/J_K^0  > 1$, the correction to
 scaling has the sign of the AF Kondo coupling $J_K$. When $J^* >0$,
 the two loop flow reinforces the tendency to flow to strong
 coupling, and comforts our identification of the strong coupling
 fixed point with the heavy Fermi liquid. When $J^* <0$ two loop
 corrections have opposite sign, revealing the presence of an
 intermediate energy scale $T_1$, where the $\beta$-function changes
 sign. When one renormalizes from high energy, the system first
 wants to form Kondo singlets, till we reach $T_1$. Further down in
 energy, for $J>J_1$ the formation of the Kondo heavy
 quasi-particles continues, while for $J< J_1$ it stops and the
 model (\ref{model}) is stable, where the three fields
 corresponding to damped spinless $\chi$-fermions, conduction
 electrons and spinons interact weakly with each other.
 The energy scale $T_1$ is seen in most of heavy Fermion
 compounds. It can be identified as the scale above which Curie
 susceptibilities are observed, or the scale at which the
 resistivity and the specific heat coefficient have a maximum.

 The precise determination of the nature of the Non Fermi Liquid
 (NFL)
 phase deserves more intensive work.
The reason why so many compounds-- with low disorder--
 show linear resistivity over a wide temperature range may lie in the
 conjecture that a damping $\Gamma (\omega ) = | \omega | \log (\omega
 )$ stabilizes our model at an intermediate energy scale.


 It has been
 advanced recently~\cite{senthil} that deconfinement of spinons
 and the emergence of a ``fractionalized'' state might be the clue
 for understanding the heavy Fermions phase diagram. We believe, however,
 that the key most probably lies in
 the idea of a competition between the formation of Kondo singlets and AF
 fluctuations, which are maximal close to a QCP.
Our study of model (\ref{model}) shows that de-confinenent of
spinons~\cite{tsvelik,chubstarykh,sachdev} is a separate issue
from Kondo singlet formation. Even though our mean-field treatment
of the Heisenberg part assumes spinon deconfinement, previous
work~\cite{wen} incites us to believe, that in $D=3$, gauge
theories being deconfining,  spinons are de-confined in the NFL
phase. The question is less clear in $D=2$.

Since in our treatment, we have assumed that the $\chi$-band is
full, we have not addressed the question of the variation of the
Fermi surface volume. In all generality, though, the chemical
potential of the $\chi$-fermion can be renormalized to zero,
leading to a reconfiguration of the charge carriers close to the
QCP.

In conclusion, we have introduced an effective theory for Kondo
lattices where the Kondo bound states are considered as true
fields, with the statistics of spinless Fermions. RG analysis of
this model signals both the formation of Kondo singlets and AF
fluctuations as logarithmic singularities at the one loop level.
This leads to a phase diagram where direct competition between
Kondo and AF is apparent, resulting in the possibility of a NFL
phase separating the AF phase from the heavy Fermi liquid. We
believe the nature of the elementary excitations at the QCP in
heavy Fermion compounds is essentially captured by (\ref{model}).

   I would like to thank K. Le Hur, I. Paul and O. Parcollet for important discussions.
    Discussions with A. Chubukov, J. Cardy, J. Custers, P. Gegenwart, M. Norman, S. Pashen, J. Rech,
    P. Simon and F. Steglich are also acknowledged.

\end{document}